\documentclass[a4paper,11pt]{article}
\usepackage[english]{babel}
\usepackage[utf8]{inputenc}
\usepackage{color}
\usepackage{amsmath,amsfonts}
\usepackage{amssymb}
\usepackage{subcaption}
\usepackage{graphicx}
\color{black}

\begin{document}
\title{Integrable generalized Heisenberg ferromagnet equations with self-consistent potentials and related Yajima-Oikawa type equations}
\author{Zhaidary  Myrzakulova$^{1,2}$\footnote{Email: zrmyrzakulova@gmail.com},  \,           Gulgassyl  Nugmanova$^{1,2}$\footnote{Email: nugmanovagn@gmail.com}, \,
Kuralay Yesmakhanova$^{1,2}$\footnote{Email: krmyrzakulova@gmail.com},\\
Nurzhan Serikbayev$^{1,2}$\footnote{Email: rmyrzakulov@gmail.com} \, and  
Ratbay Myrzakulov$^{1,2}$\footnote{Email: rmyrzakulov@gmail.com}\\
\textsl{$^{1}$Eurasian National University,
Nur-Sultan, 010008, Kazakhstan}\\
\textsl{$^{2}$Ratbay Myrzakulov Eurasian International Centre for Theoretical Physics}, \\ \textsl{Nur-Sultan, 010009, Kazakhstan}  
}
\date{}
\maketitle

\begin{abstract}
We consider some nonlinear models describing interactions of  long and short (LS) waves. Such LS models  have been derived and proposed with various motivations, which mainly come from fluid and plasma physics. In this paper, we study some of integrable LS  models, namely, the Yajima-Oikawa equation, the Newell equation, the Ma equation, the Geng-Li equation and etc.  In particular, the gauge equivalent counterparts of these integrable LS models (equations) are found. In fact, these gauge equivalents of the LS equations are integrable generalized Heisenberg ferromagnet equations (HFE) with self-consistent potentials (HFESCP). The associated Lax representations of these HFESCP are given.   We also presented  several spin-phonon equations which describe nonlinear interactions of spin and lattice subsystems in ferromagnetic materials.
\end{abstract}
%%%%%%%%%%%%%%%%%%%%%%%%%%%
\tableofcontents
\section{Introduction}

Nonlinear  models  of long wave-short wave resonant interactions, the so-called LS equations,  play important role in modern physics as well as  in modern  mathematics  \cite{yoe}-\cite{melnikov}. Some of  these models are integrable nonlinear partial differential equations in 1+1 and 2+1 dimensions (see, e.g. \cite{geng}-\cite{ustinov} and references therein). In 1+1 dimensions, such LS equations   of long wave-short wave resonant interactions, formally in general,  can be written as
\begin{eqnarray}
iq_{t}+q_{xx}+f_{1}(v, v_{x}, q, \bar{q}, ...)&=&0,\\
v_{t}+\delta(|q|^{2})_{x}&=&0, 
\end{eqnarray}
where  $q(x,t)$  represents the envelope (complex-valued) of the short wave and $v(x,t)$  represents the amplitude
(real-valued) of the long wave (potential), $\delta=const$. Here  $f_{1}$ is some function of its arguments. The set of equations (1)-(2) can be considered as the nonlinear Schr\"{o}dinger type equations  with self-consistent potentials.  It  combines all  well known integrable LS
models, namely,  the Yajima-Oikawa equation (YOE), the Newell equation (NE), the Geng-Li equation (GLE) etc.  This outcome is similar to the one that proves that the Korteweg-de
Vries  and the modified-KdV equations are just two particular cases of the Gardner equation. There  exists  another interesting class of integrable systems, namely, integrable generalized spin systems or Heisenberg ferromagnet  equations  with self-consistent potentials (GHFESCP) (see, e.g. \cite{ishimori}-\cite{s4} and references therein). Such spin systems in general can be   written  as
\begin{eqnarray}
iS_{t}+f_{2}(S,S_{x}, S_{xx}, u, u_{x}, ...)&=&0,\\
u_{t}+\nu tr(S[S_{x},S_{t}])_{x}&=&0, 
\end{eqnarray}
where the $2\times 2$ spin matrix $S$ has the form
\begin{eqnarray}
S= \begin{pmatrix}
S_{3} & S^{-}  \\
S^{+} & -S_{3}
\end{pmatrix}
\end{eqnarray} and satisfies the following condition
\begin{eqnarray}
S^{2}=I.
\end{eqnarray} 
Some time we use the following form of integrable generalized spin systems
\begin{eqnarray}
iR_{t}+f_{3}(R,R_{x}, R_{xx}, u, u_{x}, ...)=0, 
\end{eqnarray}
where in contrast to (5) and (6), $R$ is the $3\times 3$ spin matrix satisfying  the  condition
\begin{eqnarray}
R^{3}=\epsilon R, \quad (\epsilon=\pm 1).
\end{eqnarray}
In above equations, $f_{2}$ and $f_{3}$ are  some matrix functions of their arguments, $u$ is a real function (potential) and  $(\nu, \delta)$  are real constants.

The set of equations (1)-(2) is  some kind  generalizations of  the nonlinear Schr\"{o}dinger  equation (NLSE): 
\begin{eqnarray}
iq_{t} + q_{xx} +2 \nu |q|^{2}q&=&0. \label{9}
\end{eqnarray}
At the same time the set (3)-(4) and (7) are some extensions of the following Heisenberg ferromagnet equation (HFE) 
\begin{eqnarray}
iS_{t} + \frac{1}{2}[S,S_{xx}]=0. \label{10}
\end{eqnarray}
Both of the NLSE (9) and HFE (10) are integrable and admit some integrable extentions in 1+1 and 2+1 dimensions (see, e.g., \cite{rm1}-\cite{z3} and references therein). It is well-known that between the NLSE (9) and the HFE (10) takes place the  gauge and geometrical equivalence \cite{laksh}-\cite{zt1979}. The aim of this paper is the  finding  such gauge equivalence between some particular reductions of the sets of equations (1)-(2), (3)-(4)  and (7).  

This paper is organized as follows. In Section 2, the integrable Tolkynay equation (TE) is presented. The Yajima-Oikawa-Mewell equation (YONE) is considered in Section 3 and its gauge equivalence with the TE was studied in Section 4. In the next two sections (Section 5 and Section 6), the gauge equivalent counterparts of the Yajima-Oikawa equation (YOE) and the Ma equation (ME) are presented.  The relation between the TE  and NE is established in Section 7. The same problem was studied in Section 8 for the Geng-Li equation (GLE). In SEction 9, the M-XXXIV equation is investigated. In Section 10, the M-V equation and its relation with the LS equations  were considered. The nonlinear magnon-phonon equations were presented in Section 11. 
The last section is devoted to some conclusions and discussions. 

%%%%%%%%%%%%%%%%%%%%%%%%%%%%%%%%%%%%%%%%%%%%%%%%%%%%%%%%%
\section{The Tolkynay equation}
%%%%%%%%%%%%%%%%%%%%%%%%%%%%%%%%%%%%%%%%%%%%%%%%%%%%%%%%%%%%%
In this paper, in particular,  we study the following Tolkynay equation (TE) 
\begin{equation}
iR_t+[R^{2},R_{x}]_{x}=0. 
\end{equation}
Here the spin matrix $R$ has the form
\begin{equation}
R=\left(\begin{array}{ccc} R_{11} & R_{12} & R_{13} \\ R_{21} & R_{22} &  R_{23}  \\ R_{31} & R_{32}  & R_{33} \end{array}\right), 
\end{equation} 
and  satisfies the following conditions
\begin{eqnarray}
R^{3}=R, \quad tr (R)=0, \quad det(R)=0. 
\end{eqnarray}
The TE is one of integrable generalized Heisenberg ferromagnet type  equation. The  Lax representation (LR) of the TE is given by
\begin{eqnarray}
\Psi_{x}&=&U_{1}\Psi, \\ 
\Psi_{t}&=&V_{1}\Psi, 
\end{eqnarray}
where
\begin{eqnarray} 
U_1&=&i\lambda R, \\
V_1&=&-i\lambda^{2}\left(R^{2}-\frac{2}{3}I\right)-\lambda [R^{2},R_{x}].
\end{eqnarray}

%%%%%%%%%%%%%%%%%%%%%%%%%%%%%%%%%%%%%%%%%
\section{The YONE}
%%%%%%%%%%%%%%%%%%%%%%%%%%%%%%%%%%%%%%%%%%%%%%%%%%%%
One of most general integrable LS equations is the Yajima-Oikawa-Newell  equation (YONE) \cite{2109.04296}
\begin{eqnarray}
iq_{t}+q_{xx}+(i\alpha v_{x}+\alpha^{2}v^{2}-\beta v-2\alpha|q|^{2})q&=&0,\\
v_{t}-2(|q|)_{x}&=&0, 
\end{eqnarray}
where the parameters $\alpha$, $\beta$ are arbitrary real constants. These parameters may be considered as independent constants which are responsible for the long-short wave cross-interaction. This system reduces  to  the YOE for $\alpha=0, \beta=1$ and to the Newell equation as $\alpha=\sigma, \beta=0$.   The YONE (18)-(19)   is integrable. Its   LR  has the form \cite{2109.04296}
\begin{eqnarray}
\Phi_{x}&=&U_{2}\Phi, \\ 
\Phi_{t}&=&V_{2}\Phi, 
\end{eqnarray}
where
\begin{eqnarray}
U_{2}&=&i\lambda\Sigma+Q, \\ 
V_{2}&=&-\lambda^{2}B_{2}+i\lambda B_{1}+B_{0}. 
\end{eqnarray}
Here
\begin{equation}
\Sigma=\left(\begin{array}{ccc} 1 & 0 & 0 \\  0 & 0 &  0  \\ 0 & 0  & -1 \end{array}\right),\quad 
Q=\left(\begin{array}{ccc} 0 & q & iv \\ \alpha  \bar{q} & 0 &   \bar{q}  \\ 
i(\alpha^2 v-\beta) & \alpha q & 0 \end{array} \right),
\end{equation}
\begin{equation}
B_{2}=\frac{i}{3}\left(\begin{array}{ccc} 1 & 0 & 0 \\ 0 & -2 & 0  \\0 & 0 & 1 \end{array} \right),
\quad B_{1}= \left(\begin{array}{ccc} 0 & i q & 0 \\ i \alpha \bar{q} & 0 & -i \bar{q}  \\ 0 & -i\alpha q & 0 \end{array}\right),
\end{equation}
\begin{equation}
B_{0}= \left(\begin{array}{ccc} -i\alpha |q|^2 & -\alpha vq +iq_x & i|q|^2 \\ - \alpha^2 v\bar{q} +\beta  \bar{q}-i \alpha \bar{q}_{x}  & 2i \alpha |q|^2 &  -\alpha v\bar{q} -i \bar{q}_{x} \\ i \alpha^2 |q|^2& -\alpha^2 vq +\beta q+i\alpha q_{x} & -i\alpha |q|^2 \end{array}\right).
\end{equation}

%%%%%%%%%%%%%%%%%%%%%%%%%%%%%%%%%%%%%%%%%%%%%%%%%%%%%%%%%%%%%%%%%%%%%%%%%%%%

\section{Gauge equivalence between the TE and the YONE}
%%%%%%%%%%%%%%%%%%%%%%%%%%%%%%%%%%%%
In this section, we establish  the gauge equivalence between the TE (11) and the YONE (18)-(19). To this purpose, let us consider the gauge transformation 
\begin{eqnarray} 
\Psi=g^{-1}\Phi, \quad g=\Phi_{\lambda=0}.  
\end{eqnarray}
We obtain 
\begin{eqnarray} 
\Psi_{x}&=&U_{1}\Psi, \\
\Psi_{t}&=&V_{1}\Psi,  
\end{eqnarray}
where 
\begin{equation} 
U_{1}=g^{-1}(U_{2}-g_{x}g^{-1})g, \quad V_{1}=g^{-1}(V_{2}-g_{t}g^{-1})g.
\end{equation}
As results, we have
\begin{equation} 
U_1=i\lambda R, \quad      V_1=-\lambda^{2}g^{-1}B_{2}g+i\lambda g^{-1}B_{1}g,
\end{equation}
where
\begin{equation} 
 R=g^{-1}\Sigma g.
\end{equation}
After some calculations, we obtain
\begin{equation} 
 Q\Sigma+\Sigma Q= \left(\begin{array}{ccc} 0 & q & 0 \\ \alpha \bar{q} & 0 & -\bar{q}  \\0 & -\alpha q & 0 \end{array} \right)=-iB_{1},
\end{equation}
or
\begin{equation} 
 \grave{g^{-1}Qgg^{-1}\Sigma g+g^{-1}\Sigma g g^{-1}Qg}= g^{-1}\left(\begin{array}{ccc} 0 & q & 0 \\ \alpha \bar{q} & 0 & -\bar{q}  \\0 & -\alpha q & 0 \end{array} \right)=-ig^{-1}B_{1}g=g^{-1}QgR+R g^{-1}Qg.
\end{equation}
On the other hand, we obtain
\begin{equation} 
 R_{x}=g^{-1}[\Sigma, Q]g= g^{-1}\left(\begin{array}{ccc} 0 & q & 2iv \\ -\alpha \bar{q} & 0 & \bar{q}  \\
-2i(\alpha^{2}v-\beta) & -\alpha q & 0 \end{array} \right)g
\end{equation}
Hence we get
\begin{equation} 
 [R,R_{x}]= g^{-1}\left(\begin{array}{ccc} 0 & q & 4iv \\ \alpha \bar{q} & 0 & \bar{q}  \\
4i(\alpha^{2}v-\beta) & \alpha q & 0 \end{array} \right)g=g^{-1}Qg+3ig^{-1}\left(\begin{array}{ccc} 0 & 0 & v \\ 0 & 0 & 0  \\
(\alpha^{2}v-\beta) & 0 & 0 \end{array} \right)g
\end{equation}
so that we  have
\begin{equation} 
 g^{-1}Qg=[R,R_{x}]-3ig^{-1}\left(\begin{array}{ccc} 0 & 0 & v \\ 0 & 0 & 0  \\
(\alpha^{2}v-\beta) & 0 & 0 \end{array} \right)g
\end{equation}
Taking into account the formula
\begin{equation} 
 \Sigma \left(\begin{array}{ccc} 0 & 0 & v \\ 0 & 0 & 0  \\
(\alpha^{2}v-\beta) & 0 & 0 \end{array} \right)+\left(\begin{array}{ccc} 0 & 0 & v \\ 0 & 0 & 0  \\
(\alpha^{2}v-\beta) & 0 & 0 \end{array} \right) \Sigma=0
\end{equation}
we obtain
\begin{equation} 
 g^{-1}B_{1}g=i(g^{-1}Qgg^{-1}\Sigma g+g^{-1}\Sigma g g^{-1}Qg)=i([R,R_{x}]R+R[R,R_{x}])=i[R^{2},R_{x}]
\end{equation}
Thus, we have shown that
\begin{eqnarray} 
g^{-1}B_{2}g&=&-\frac{2i}{3}I+iR^{2}, \\
g^{-1}B_{1}g&=&i[R^{2},R_{x}].  
\end{eqnarray}
Finally, we have the following Lax pair for the TE (11)
\begin{eqnarray} 
U_1&=&i\lambda R, \\
V_1&=&-i\lambda^{2}(R^{2}-\frac{2}{3}I)-\lambda [R^{2},R_{x}].
\end{eqnarray}
Let us present some useful formulas
\begin{eqnarray} 
tr(R_{x}^{2})=-4\alpha|q|^{2}+8v(\alpha^{2}v-\beta), \quad det(R_{x})=2i\beta|q|^{2}.
\end{eqnarray}
As integrable equation, the YONE admits the infinity number of integrals of motion. For example,  here we present the following integral of motion for the YONE (18)-(19):
\begin{eqnarray} 
P=\int Jdx,
\end{eqnarray}
where $a, b$ are some constants and
\begin{eqnarray} 
J=4a[\alpha(5\beta-1)|q|^{2}+2v(\alpha^{2}v-\beta)]=atr(S_{x}^{2})+bdet(S_{x}), \quad b=\frac{10\alpha a}{i\beta}, \quad a=const, \quad b=const.
\end{eqnarray}
 In fact, the quantity $J$ satisfies the following conservation equation
\begin{eqnarray} 
J_{t}=16a[i\alpha(\bar{q}_{x}q-\bar{q}q_{x})-\beta|q|^{2}+2\alpha^{2}v|q|^{2}]_{x}
\end{eqnarray}
so that 
\begin{eqnarray} 
P_{t}=\left(\int Jdx\right)_{t} =0
\end{eqnarray}
for the boundary conditions 
\begin{eqnarray} 
\lim q(x,t) \rightarrow 0, \quad \lim v(x,t)\rightarrow 0 \quad as \quad  x\rightarrow \pm \infty.
\end{eqnarray}

%%%%%%%%%%%%%%%%%%%%%%%%%%%%%%%%%%%%%%%%%%%%%%%%%%%%%%%%%%%%
\section{Gauge equivalence between the YOE and the MM-IIE}
%%%%%%%%%%%%%%%%%%%%%%%%%%%%%%%%%%%%%%%%%%%%%%%%%%%%%%%%%%%%

The  first example of integrable  long-short waves interactions models is the following Yajima-Oikawa equation (YOE) \cite{yoe}
\begin{eqnarray} 
iq_t +\frac{1}{2}q_{xx}-uq=0,  \\
u_t+u_x+|q|_{x}^{2}=0.
\end{eqnarray}
Its Lax representation reads as
\begin{eqnarray}
U_{3}&=&A_0+2i\lambda \Sigma+(2\lambda)^{-1}A_{-1},\\
V_{3}&=&-U+2i\lambda^2 A_{1}^2 +\lambda B_1 +B_0 +i(4\lambda)^{-1} 
\left(\begin{array}{ccc}  |\Phi|^2 & 0 & |\Phi|^2 \\ \Phi_x &  0 & \Phi_x \\ -|\Phi|^2  & 0 & -|\Phi|^2 \end{array}\right),
\end{eqnarray}
where
\begin{eqnarray}
\Sigma&=& \left(\begin{array}{ccc} 1 & 0 & 0 \\ 0 & 0 & 0 \\ 0 & 0 & -1 \end{array}\right), 
A_0=\left(\begin{array}{ccc} 0 & -\Phi^* & 0 \\ 0 & 0 & 0 \\ 0 & \Phi^* & 0 \end{array}\right),
A_{-1}=\left(\begin{array}{ccc} -ni & 0 & -ni \\ \Phi & 0 & \Phi \\ ni & 0 & ni \end{array}\right),\\
B_1&=&\left(\begin{array}{ccc} 0 & -\Phi^* & 0 \\ 0 & 0 & 0 \\ 0 & \Phi^* & 0 \end{array}\right),
B_0=\left(\begin{array}{ccc} 0 & \frac{i}{2} \Phi_x^* +\Phi^* & 0 \\ \frac{1}{2} \Phi & 0 & - \frac{1}{2} \Phi  \\ 0 & - \frac{i}{2} \Phi_x^*- \Phi^* & 0 \end{array}\right), \\
B_{-1}&=&\left(\begin{array}{ccc} |\Phi|^2 & 0 & |\Phi|^2 \\ \Phi_x & 0 & \Phi_x \\ -|\Phi|^2 & 0 & -|\Phi|^2 \end{array}\right), \quad 
\Phi=q e^{i(\frac{t}{2}-x)}.
\end{eqnarray}
Consider a gauge transformation $\phi=g\bar\phi, \quad g(x, \ t, \ \lambda_0)=\phi(x,\ t, \ \lambda) |_{\lambda=\lambda_0} \ \subset GL(\xi, \ C)$ as $\lambda=\lambda_0$, where
\begin{equation} \label{r1.11}
g_x=U_{3}(x,t,\lambda_0)g,  \quad    g_t=V_{3}(x,t,\lambda_0)g.
\end{equation}
Then
\begin{equation} 
U_{4}=g^{-1}U_{3}g - g^{-1}g_x, \quad 
V_{4}=g^{-1}V_{3}g-g^{-1}g_t.
\end{equation}
We obtain
\begin{eqnarray} 
U_{4}&=& 2i(\lambda - \lambda_0)g^{-1}A_1 g - \frac{\lambda - \lambda_0}{4 \lambda_0 \lambda} g^{-1} A_{-1} g,\\
V_{4} &=& 2i(\lambda^2 - \lambda_0^2) g^{-1} A^2_1 g - U_{4} - \frac{i(\lambda - \lambda_0)}{4 \lambda_0 \lambda} g^{-1} B_{-1} g + (\lambda - \lambda_0) B_1.
\end{eqnarray}

Now let us introduce two new matrices $R$ and $\sigma$  as
\begin{equation} \label{r1.16}
R = g^{-1}\Sigma g , \quad 
\sigma=g^{-1} A_{-1} g, \quad R^{3}=R. 
\end{equation}
Then 
\begin{eqnarray} 
 U_{4}&=& 2i(\lambda - \lambda_0)[R +(4 \pi \lambda_0 \lambda)^{-1}\sigma],\\
V_{4}&=& - \tilde U + (\lambda - \lambda_0)\{2i(\lambda+ \lambda_0) R^{2}+R (R^2)_x + \lambda^{-1} [i(4 \lambda_0)^{-1}[\sigma,R^2] + \frac{1}{2}[\sigma,R^2]R]\}.
\end{eqnarray}
This Lax pair gives the following Makhankov-Myrzakulov-II equation (MM-IIE) \cite{makhankov}
\begin{eqnarray} 
R_t + R_x + (\frac{i}{2} R R_x^2 -2 \lambda_0 R_x^2)_x + (4i \lambda_0)^{-1}h&=&0,\\
\sigma_t + \sigma_x + (\frac{1}{2}[\sigma, R^2]_x - i \lambda_0 [\sigma, R^2])_x + i \lambda_0 h&=&0,
\end{eqnarray}
where
\begin{eqnarray}
h&=&[\sigma(R_x - 2 i \lambda_0I), R^2]+ ([R^2, \sigma]R)_x,\\
tr(U_{4}) &=& tr(V_{4}) = tr(U_{3})=tr(V_{3})=tr(R)=tr(\sigma)=0.
\end{eqnarray}

%%%%%%%%%%%%%%%%%%%%%%%%%%%%%%%%%%%%%%%%%%%%%%%%%%%%%%%%
\section{The MM-IE and its relation with the  Ma equation}
%%%%%%%%%%%%%%%%%%%%%%%%%%%%%%%%%%%%%%%%%%%%%%%
Let us consider the following Makhankov-Myrzakulov-I equation (MM-IE) \cite{makhankov}
\begin{equation} 
i R_t + 2[R, R_{xx}] - 4(R_x R)_x=0,
\end{equation}
where the spin matrix $R$ satisfies the following condition
\begin{equation} 
R^{3}=R. 
\end{equation}
The MM-IE is integrable. The corresponding LR has the form 
\begin{eqnarray} 
\Psi_{x}&=&U_{5}\Psi, \\ 
\Psi_{t}&=&V_{5}\Psi, 
\end{eqnarray}
where
\begin{eqnarray} 
 U_5&=&i \lambda R, \\ 
 V_5&=&2i \lambda^2 R^2 + 2 \lambda(R^2)_x. 
\end{eqnarray}

Note that the MM-IE is gauge equivalent to the following Ma equation (ME) \cite{makhankov}
\begin{eqnarray} 
iq_t - 2q_{xx}+ 2uq=0, \\
u_t+|q|_x^2=0,
\end{eqnarray}
In fact, 
after some simple scale transformations, the ME takes the form of the YOE (50)-(51).  In contrast to the YOE, for the ME the LR takes the more simple form \cite{ma}
\begin{eqnarray} 
\Phi_x&=&U_{6}\Phi, \\
\Phi_t&=&V_{6}\Phi,
\end{eqnarray}
where
\begin{eqnarray} 
U_6=C_0 + i \lambda \Sigma, \quad V_6=D_0 + \lambda D_1 + 2i \lambda^2 D_2.
\end{eqnarray}  
Here
\begin{eqnarray}
C_0&=& \left(\begin{array}{ccc} 0 & \frac{E}{2} & in \\ 0 & 0 & \frac{E^*}{2} \\ -i & 0 & 0 \end{array}\right), \quad 
\Sigma= \left(\begin{array}{ccc} 1 & 0 & 0 \\ 0 & 0 & 0 \\ 0 & 0 & -1 \end{array}\right),\\
D_0&=& \left(\begin{array}{ccc} 0 & -iE_x & - \frac{i}{2}|E|^2 \\ -E^* & 0 & iE_x^* \\ 0 & -E & 0 \end{array}\right), \quad
D_1= \left(\begin{array}{ccc} 0 & E & 0 \\ 0 & 0 & -E^* \\ 0 & 0 & 0 \end{array}\right), \quad D_2=\Sigma^2.
\end{eqnarray}

As we mentioned, the MM-IE and the ME is gauge equivalent to each other. To prove this, let us consider the gauge transformation $\Psi=\omega^{-1}\Phi$, where  $\omega=\Phi(x, t,  \lambda)|_{\lambda=0}$. Then we have 
\begin{eqnarray}
 \omega_x&=&U_{60} \omega, \\ 
\omega_t&=&V_{60} \omega, 
\end{eqnarray}
where
$$
U_{60}=U_1 \ |_{\lambda=0}, \quad V_{60}=V_1 \ |_{\lambda=0}.
$$
As result, we get 
\begin{equation} \label{r1.26}
 U_5=\omega^{-1}U_6 \omega - \omega^{-1} \omega_x, \quad 
 V_5=\omega^{-1}V_6 \omega - \omega^{-1} \omega_t,
\end{equation}
or
$$
 U_6=i \lambda \omega^{-1} \Sigma\omega, \quad  V_6= \lambda \omega^{-1} D_1 \omega + 2 i \lambda^2 \omega^{-1} D_2 \omega.
$$
After some calculation we obtain
$$
\omega^{-1} D_1 \omega= 2 (R^2)_x, \quad 
\omega^{-1}D_2 \omega=R^2,
$$
where
\begin{equation} \label{r1.27}
R= \omega^{-1}\Sigma \omega.
\end{equation}
The matrix function $R$ satisfies the following conditions
\begin{equation} \label{r1.30}
tr(R) =0,\quad 
\det(R) =\frac{i}{2} |E|^2, \quad \det(R_{x})= \det(\omega^{-1}[C_1,C_0] \omega)=\frac{i}{2} |E|^2.
\end{equation}
It is not difficult to verify that the matrix function $R$ satisfies the MM-IE (68). This is proves that the MM-IE (68) and the ME (74)-(75)) is gauge equivalent to each other.

%%%%%%%%%%%%%%%%%%%%%%%%%%%%%%%%%%%%%%%%
\section{The TE and its relation with the Newell equation}
%%%%%%%%%%%%%%%%%%%%%%%%%%%%%%%%%%%%%%%%
The TE has the form
\begin{equation}
iR_t+[R^{2},R_{x}]_{x}=0, 
\end{equation}
where the spin matrix $R$ has the form
\begin{equation}
R=\left(\begin{array}{ccc} R_{11} & R_{12} & R_{13} \\ R_{21} & R_{22} &  R_{23}  \\ R_{31} & R_{32}  & R_{33} \end{array}\right)=\left(\begin{array}{ccc} R_{11} & R_{12} & iv \\ -\sigma \bar{R}_{12} & 0 &  \sigma\bar{R}_{12} \\ -iv & -R_{12}  & -R_{11} \end{array}\right)=\left(\begin{array}{ccc} R_{3} & R^{-} & iv \\ -\sigma R^{+} & 0 &  \sigma R^{+} \\ -iv & -R^{-}  & -R_{3} \end{array}\right), 
\end{equation} 
and  satisfies the following conditions
\begin{eqnarray}
R^{3}=R, \quad tr (R)=0, \quad det(R)=0. 
\end{eqnarray}
Here $\sigma=\pm 1$,  $R_{12}=R^{-}=R_{1}-iR_{2}$ is a complex function and $R_{11}=R_{3}, R_{1}, R_{2},  v$ are real functions. Let us calculate some useful expressions
\begin{equation}
R^{2}=\left(\begin{array}{ccc} R_{11}^{2}-\sigma |R_{12}|^{2}+v^{2} & (R_{11}-iv)R_{12} & \sigma|R_{12}|^{2} \\ 
-(R_{11}+iv)\sigma \bar{R}_{12} & -2\sigma|R_{12}|^{2} &  -(R_{11}+iv)\sigma \bar{R}_{12}  \\ 
\sigma|R_{12}|^{2} & (R_{11}-iv) R_{12}  & R_{11}^{2}-\sigma|R_{12}|^{2}+v^{2} \end{array}\right), 
\end{equation} 
\begin{equation}
R^{3}=R^{2}R= R=(R_{11}^{2}-2\sigma|R_{12}|^{2}+v^{2})\left(\begin{array}{ccc} R_{11} & R_{12} & iv \\ -\sigma \bar{R}_{12} & 0 &  \sigma\bar{R}_{12} \\ -iv & -R_{12}  & -R_{11} \end{array}\right)=(R_{11}^{2}-2\sigma|R_{12}|^{2}+v^{2})R. 
\end{equation}
Hence we obtain  the following condition
\begin{eqnarray}
R_{11}^{2}-2\sigma|R_{12}|^{2}+v^{2}=R_{3}^{2}-2\sigma|R^{+}|^{2}+v^{2}=R_{3}^{2}-2\sigma(R_{1}^{2}+R_{2}^{2})+v^{2}=1. 
\end{eqnarray}
The TE is one of integrable generalized Heisenberg ferromagnet type  equation. The  LR of the TE is given by
\begin{eqnarray}
\Psi_{x}&=&U_{7}\Psi, \\ 
\Psi_{t}&=&V_{7}\Psi, 
\end{eqnarray}
where
\begin{eqnarray} 
U_7&=&i\lambda R, \\
V_7&=&-i\lambda^{2}\left(R^{2}-\frac{2}{3}I\right)-\lambda [R^{2},R_{x}].
\end{eqnarray}

It is not difficult to verify that the gauge equivalent of TE (86) is the Newell equation (NE) which reads as \cite{newell}
\begin{eqnarray}
i q_t +q_{xx} +( i u_x + u^2 -2\sigma |q|^2)q&=&0\\
u_t-2\sigma (|q|^2)_x&=&0,
\end{eqnarray}
where $\sigma=\pm 1$. Note that in addition to a long wave-short wave  coupling, the short wave has the same self-interaction as the NLS equation (9). The NE  is  integrable. Its   Lax representation   looks like \cite{newell}, \cite{ling}
\begin{eqnarray}
\Phi_{x}&=&U_{8}\Phi, \\ 
\Phi_{t}&=&V_{8}\Phi, 
\end{eqnarray}
where
\begin{eqnarray}
U_{8}=\left(\begin{array}{ccc} i\lambda & q & iv \\  
\sigma \bar{q} & 0 &\sigma \bar{q}   \\ iv & q  & -i\lambda \end{array}\right),
V_{8}=\left(\begin{array}{ccc} -\frac{1}{3}i\lambda^{2}-i\sigma|q|^{2} & -\lambda q+iq_{x}-vq & i\sigma |q|^{2}
 \\ -\sigma(\lambda \bar{q}+\bar{q}_{x}+v\bar{q}) & \frac{2}{3}i\lambda^{2}+2i\sigma|q|^{2} &   \sigma(\lambda \bar{q}-i\bar{q}_{x}-v\bar{q})  \\ 
i\sigma|q|^{2} & \lambda q+iq_{x}-vq & -\frac{1}{3}i\lambda^{2}-i\sigma|q|^{2} \end{array}\right).
\end{eqnarray}
 It is not difficult to verify that these matrices satisfy the following conditions
\begin{eqnarray}
U^{+}(\lambda)=-AU(\bar{\lambda})A, \ V^{+}(\lambda)=-AV(\bar{\lambda})A,\, U^{+}(\lambda)=-BV(-\lambda)B,\, U^{+}(\lambda)=-BV(-\lambda)B,
\end{eqnarray}
where
\begin{eqnarray}
A=\left(\begin{array}{ccc} 1 & 0 & 0 \\  
0 & -\sigma &0   \\ 0& 0  & 1\end{array}\right),\quad 
B=\left(\begin{array}{ccc} 0 & 0 & 1
 \\ 0 & 1 &   0  \\ 
1 & 0 & 0 \end{array} \right).
\end{eqnarray}
%%%%%%%%%%%%%%%%%%%%%%%%%%%%%%%%%%%%%%%%%%%%%%%%%%%%%%%%%%%%%

%%%%%%%%%%%%%%%%%%%%%%%%%%%%%%%%%%%%%%
\section{Geng-Li equation}
%%%%%%%%%%%%%%%%%%%%%%%%%%%%%%%%%%%%%
%%%%%%%%%%%%%%%%%%%%%%%%%%%%%%%%%%%%%%%%%
Again we consider TE 
\begin{equation}
iR_t+[R^{2},R_{x}]_{x}=0. 
\end{equation}
In contrast to the previous cases, now we assume that  the spin matrix $R$ satisfies the conditions
\begin{eqnarray}
R^{3}=-R, \quad tr (R)=0, \quad det(R)=0. 
\end{eqnarray}
Then the LR for the TE (103) takes the form
\begin{eqnarray}
\Psi_{x}&=&U_{9}\Psi, \\ 
\Psi_{t}&=&V_{9}\Psi, 
\end{eqnarray}
where
\begin{eqnarray} 
U_9&=&\lambda R, \\
V_9&=&-i\lambda^{2}R^{2}+i\lambda [R^{2},R_{x}].
\end{eqnarray}
Now we assume that  the matrix $R$ can be written as 
\begin{equation}
R=g^{-1}Jg,
\end{equation} 
where
\begin{equation}
J=\left(\begin{array}{ccc} 0 & -1 & 0 \\ 1 & 0 &  0 \\ 0& 0  & 0 \end{array}\right). 
\end{equation}
Consider the gauge transformation
\begin{eqnarray}
\Phi=g\Psi. 
\end{eqnarray}
Hence and from the equations (105)-(106) follows that the new matrix function $\Phi$ satisfies the equations 
\begin{eqnarray}
\Phi_{x}&=&U_{8}\Phi, \\ 
\Phi_{t}&=&V_{8}\Phi, 
\end{eqnarray}
where
\begin{eqnarray}
U_{8}=\left(\begin{array}{ccc} iv & -\lambda & 0 \\  
\lambda & 0 & -\bar{q}   \\ 0 & q  & 0 \end{array}\right),\quad 
V_{8}=\left(\begin{array}{ccc} i\lambda^{2}-2i|q|^{2} & 0 & -i\lambda \bar{q}
 \\ 0 & i\lambda^{2}-i|q|^{2} &   i\bar{q}_{x}+v\bar{q}  \\ 
-i\lambda q & iq_{x}-vq & i|q|^{2} \end{array} \right).
\end{eqnarray}
The compatibility condition 
\begin{eqnarray}
U_{8t}-V_{8x}+[U_{8},V_{8}]=0 
\end{eqnarray}
gives the Geng-Li equation (GLE)  \cite{geng}
\begin{eqnarray}
iq_{t}+q_{xx}+2|q|^{2}q+i(vq)_{x}&=&0,\\
v_{t}+2(|q|^{2})_{x}&=&0. 
\end{eqnarray}
This proves that the TE (103) with the conditions (104) and LR (105)-(106) is gauge equivalent to the GLE (116)-(117). 
%%%%%%%%%%%%%%%%%%%%%%%%%%%%%%%%%%%%%%%%%%%%%%%%%%%%%%%%%%%%%%%%%%%%%
\section{The M-XXXIV equation}
%%%%%%%%%%%%%%%%%%%%%%%%%%%%%%%%%%%%
One of integrable HFE with self-consistent potentials (HFESCP) is the following Myrzakulov-XXXIV (M-XXXIV) equation  
\begin{eqnarray} 
{\bf S}_{t}+{\bf S}\wedge{\bf S}_{xx}-u{\bf S}&=&0, \\
u_{t}+\frac{1}{2}({\bf S}_{x}^{2})_{x}&=&0,  
\end{eqnarray}
where ${\bf S}=(S_{1},S_{2},S_{3})$ is the unit spin vector that is ${\bf S}^{2}=S_{1}^{2}+S_{2}^{2}+S_{3}^{2}=1$ and  $u$ is a real function (potential). The M-XXXIV equation  is integrable. The corresponding LR has the form 
\begin{eqnarray} 
\alpha \Psi_y&=&\frac{1}{2}[S+I]\Psi_x, \\
 \Psi_t&=&\frac{i}{2}[S+(2b+1)I]\Psi_{xx}+\frac{i}{2}W\Phi_x, 
\end{eqnarray}
where $W=W_{1}+W_{2}$ and 
\begin{eqnarray} 
 W_1&=&(2b+1)E+(2b+\frac{1}{2})SS_x+(2b+1)FS,\\
 W_2&=&FI+\frac{1}{2}S_x+ES+\alpha SS_y,\quad S^{\pm}=S_{1}\pm iS_{2}, \\
E&=& -\frac{i}{2\alpha} u_x,\quad   F = \frac{i}{2}\left(\frac{u_{x}}{\alpha} -
2u_{y}\right), \quad S=\begin{pmatrix}
S_{3} & S^{-} \\
S^{+} & -S_{3}
\end{pmatrix}.
\end{eqnarray}
In fact, the compatibility condition $\Psi_{yt}=\Psi_{ty}$ gives the following set of equations
\begin{eqnarray}
 iS_t+\frac{1}{2}[S,S_{\xi\xi}]-iwS_{\xi}&=&0,\\
w_{\eta}-\frac{1}{4i}tr(S[S_{\xi},S_{\eta}])&=&0,
\end{eqnarray}
where
\begin{eqnarray}
\xi = x+\frac{1}{\alpha}y,\quad 
\eta = -x,\quad  w=u_{\xi}.
\end{eqnarray}
Hence after the simple transformation $\eta=t, w\rightarrow u, \xi \rightarrow x$, we obtain 
\begin{eqnarray}
 iS_t+\frac{1}{2}[S,S_{xx}]-iuS_{x}&=&0,\\
u_{t}-\frac{1}{4i}tr(S[S_{x},S_{t}])&=&0,
\end{eqnarray}
The following  equations  are correct:
\begin{eqnarray}
 [S_{x}, S_t]=-i(S^{2}_{x})_{x}S, \quad S[S_{x}, S_t]=-i(S^{2}_{x})_{x}I,\quad S^{2}_{x}={\bf S}_{x}^{2} I, \quad S[S_{x}, S_t]=-i({\bf S}^{2}_{x})_{x}I,
\end{eqnarray}
and 
\begin{eqnarray}
 tr(S[S_{x}, S_t])=-2i({\bf S}^{2}_{x})_{x}.
\end{eqnarray}
Hence the M-XXXIV equation  (128)-(129) can be written as 
\begin{eqnarray}
 iS_t+\frac{1}{2}[S,S_{xx}]-iuS_{x}&=&0,\\
u_{t}+\frac{1}{2}({\bf S}^{2}_{x})_{x}&=&0.
\end{eqnarray}
Let us now find the equation which is gauge equivalent to the M-XXXIV
equation (132)-(133). To this end, we consider the following tranformation
\begin{eqnarray}
\Phi = g \Psi,
\end{eqnarray}
where $\Psi$ is the matrix solution of linear problem (120)-(121), $\Phi$ and
$g$ are a temporally unknown matrix functions. Substituting (134) into (120)-(121) after some calculations we get
\begin{eqnarray}
 \alpha \Phi_y &=&B_{1}\Phi_x + B_{0}\Phi, \\
 \Phi_t&=&iC_{2}\Phi_{xx}+C_{1}\Phi_x+C_{0}\Phi, 
\end{eqnarray}
with
\begin{eqnarray}
B_{1}&=& \begin{pmatrix}
1 & 0  \\
0 & 0
\end{pmatrix}, \quad B_{0}= \begin{pmatrix}
0 & q  \\
r & 0
\end{pmatrix},\\
C_{2}&=& \begin{pmatrix}
b+1 & 0 \\
0   & b
\end{pmatrix}, \quad C_{1}= \begin{pmatrix}
0   &  iq \\
ir  &  0
\end{pmatrix},\quad 
C_{0}= \begin{pmatrix}
c_{11}  &  c_{12} \\
c_{21}  &  c_{22}
\end{pmatrix},\\
c_{12}&=&i(2b+1)q_{x}+i\alpha q_{y},\quad c_{21}=-2ib r_{x}-i\alpha r_{y}.
\end{eqnarray}
Here $c_{jj}$ satisfy the following system of equations
\begin{eqnarray}
 c_{11x}- \alpha c_{11y} =iqr_{x}+r c_{12} -q c_{21},\quad
\alpha c_{22y}=-irq_{x}+rc_{12}-qc_{21}.
\end{eqnarray}

The compatibility condition of equations(135)-(136) gives the following
(2+1)-dimensional nonlinear Schr\"odinger  equation:
\begin{eqnarray}
iq_{t}+q_{\xi \xi}+vq&=&0, \\
ir_{t}-r_{\xi\xi}-vr&=&0, \\
v_{\eta} +2(r q)_{\xi}&=&0, 
\end{eqnarray}
or after $\eta\rightarrow t$ we have 
\begin{eqnarray}
iq_{t}+q_{xx}+vq&=&0, \\
ir_{t}-r_{xx}-vr&=&0, \\
v_{t} +2(rq)_{x}&=&0. 
\end{eqnarray}
It coincide with the  ME (74)-(75). Thus we have presented  the new LR for the ME or for the YOE. Consequently we found the new form of the gauge equivalent counterpart of the ME and/or the YOE, namely, the M-XXXIV equation. 

%%%%%%%%%%%%%%%%%%%%%%%%%%%%%%%%%%%%%%%%%%%%%%%
\section{The M-V equation}
%%%%%%%%%%%%%%%%%%%%%%%%%%%%%%%%%%%%%%%%%%%%%%%%%
%%%%%%%%%%%%%%%%%%%%%%%%%%%%%%%%%%%%%%%%%%%%%%%%%%%%%%%%%
Our next example of integrable generalized HFE is the so-called Myrzakulov-V (M-V) equation. The M-V equation   reads as
\begin{equation}
iR_t+2[R, R_{y}]_{x}+3(R^{2}R_{y}R)_{x}=0
\end{equation}
or
\begin{equation}
iR_t+\frac{1}{2}[R, R_{y}]_{x}+\frac{3}{2}[R^{2},(R^{2})_{y}]_{x}=0,
\end{equation}
where the spin matrix $R$   satisfies the following conditions
\begin{eqnarray}
R^{3}=R, \quad tr (R)=0, \quad det(R)=0. 
\end{eqnarray}
The M-V equation (148)  is the (2+1)-dimensional integrable equation. Its LR looks like 
\begin{eqnarray}
\Psi_{x}&=&U_{1}\Psi, \\ 
\Psi_{t}&=&V_{1}\Psi, 
\end{eqnarray}
where
\begin{eqnarray} 
U_1&=&-i\lambda R, \\
V_1&=&-2i\lambda^{2}R+\frac{\lambda }{2}\left([R,R_{y}]+3[R^{2}, (R^{2})_{y}]\right).
\end{eqnarray}
To find its gauge equivalent equation, let us consider the transformation
\begin{equation}
R=\Phi^{-1}\Sigma \Phi,
\end{equation}
where $\Sigma=diag(1, 0, -1)$. Let us we assume that $\Phi$ satisfies the following equations
\begin{equation}
\Phi_{x}=-i\lambda \Sigma +Q, \quad \Phi_{t}=(\mu_{2}\lambda^{2}+\mu_{1}\lambda+\mu_{0}) \Phi_{y}+V\Phi,
\end{equation}
where the matrix $Q$  is given,   $V$ is unknown matrix and $\mu_{j}=consts$. The compatibility condition $\Phi_{xt}=\Phi_{tx}$ gives the following two equations
\begin{eqnarray}
Q_{t}-V_{x}+[U,V]-(\mu_{2}\lambda^{2}+\mu_{1}\lambda+\mu_{0})Q_{y} =0 
\end{eqnarray}
and 
\begin{eqnarray}
\lambda_{t}-(\mu_{2}\lambda^{2}+\mu_{1}\lambda+\mu_{0})\lambda_{y}=0. 
\end{eqnarray}
The equation (156) is the desired nonlinear Schr\"{o}dinger type equation coupled with the equation for the potential $v(x,t)$.  At the same time, the equation (157) tells us that in this case, we have the nonisospectral problem, where $\lambda=\lambda(y,t)$.

%%%%%%%%%%%%%%%%%%%%%%%%%%%%%%%%%%%%%%%%%%%%%%%%%%%%%%%%%%%
\section{Magnon-phonon models as HFE with self-consistent potentials}
%%%%%%%%%%%%%%%%%%%%%%%%%%%%%%%%%%%%%%%%%%%%%%%%%%%%%%%%%%%%%%%%%%%%%%%%

In this section we want to present some HFE with self-consistent potentials (HFESCP), namely, the so-called magnon-phonon systems or spin-phonon systems. These models describe nonlinear interactions between the spin waves and the lattice waves. Some of these models have the different classes  of nonlinear solutions:  solitons, kinks, breathers and so on.  We continue to study the key properties of integrable spin-phonon nonlinear systems serving to mimicry some
dynamical features of physically motivated but non-integrable
spin-phonon nonlinear models. Some of these models are integrable but others are nonintegrable. In particular, the Kuralay-I equation (166)-(167), the Shynaray-I equation (168)-(170), the Zhaidary-I equation (171)-(173) are integrable, admit Lax representations  and so on.   Here some of these spin-phonon or magnon-phonon models.

M-XXXIII equation:
\begin{eqnarray} 
2iS_t=[S,S_{xx}]+2iuS_x, \\
u_t+u_x+\alpha(u^2)_x+\beta u_{xxx}+\frac{\lambda}{4}tr(S^2_x)_x=0.
\end{eqnarray}

M-XXXIV equation:
\begin{eqnarray} 
{\bf S}_t = {\bf S} \wedge {\bf S}_{xx}+u{\bf S}_{x},\\
u_{t}+4\eta({\bf S}^{2}_{x})_{x}=0.
\end{eqnarray}

M-XXXV equation: 
\begin{eqnarray} 
2iS_t=[S,S_{xx}]+2iuS_x,\\
\rho u_{tt}=\nu^2_0u_{xx}+\alpha(u^2)_{xx}+\beta u_{xxxx}+\frac{\lambda}{4}tr(S^2_x)_x.
\end{eqnarray}

M-XXXVI equation:
\begin{eqnarray} 
2iS_t=[S,S_{xx}]+2iuS_x,\\
\rho u_{tt}=\nu^2_0u_{xx}+\frac{\lambda}{4}tr(S^2_x)_x.
\end{eqnarray}

Kuralay equation:
\begin{eqnarray}
{\bf S}_{t}-{\bf S}\wedge {\bf S}_{xt}-u{\bf S}_{x}&=&0,\\
u_{x}+\frac{1}{2}({\bf S}_{x}^{2})_{t}&=&0.
\end{eqnarray}

Shynaray equation: 
\begin{eqnarray}
{\bf S}_{t}-{\bf S}\wedge {\bf S}_{xt}-u{\bf S}_{x}-2cl^{2}S_{t}-4c(w{\bf S})_{x}&=&0,\\
u_{x}+\frac{1}{2}({\bf S}_{x}^{2})_{t}&=&0.\\
w_{x}+\frac{1}{16l^{2}c^{2}}({\bf S}_{x}^{2})_{t}&=&0.
\end{eqnarray}

Zhaidary equation:
\begin{eqnarray}
{\bf S}_{t}-{\bf S}\wedge {\bf S}_{xt}-u{\bf S}_{x}-2l(cl+d)S_{t}-4c(w{\bf S})_{x}&=&0,\\
u_{x}+\frac{1}{2}({\bf S}_{x}^{2})_{t}&=&0.\\
w_{x}+\frac{1}{4(2lc+d)^{2}}({\bf S}_{x}^{2})_{t}&=&0
\end{eqnarray}

M-XXXVII equation:
\begin{eqnarray} 
2iS_t=[S,S_{xxxx}]+2\left\{((1+\mu)\vec{S}^2_x-u+m)[S,S_x]\right\}_x,\\
u_{t}+u_{x}+\alpha(u^2)_{x}+\beta u_{xxx}+\lambda(\vec{S}^2_x)_x=0.
\end{eqnarray}

M-XXXVIII equation:
\begin{eqnarray} 
2iS_t=[S,S_{xxxx}]+2\left\{((1+\mu)\vec{S}^2_x-u+m)[S,S_x]\right\}_x, \\ 
u_{t}+u_{x}+\lambda(\vec{S}^2_x)_x=0.
\end{eqnarray}

M-XXXIX equation:
\begin{eqnarray} 
2iS_t=[S,S_{xxxx}]+2\left\{((1+\mu)\vec{S}^2_x-u+m)[S,S_x]\right\}_x,\\
\rho u_{tt}=\nu^2_0u_{xx}+\alpha(u^2)_{xx}+\beta u_{xxxx}+\lambda(\vec{S}^2_x)_{xx}.
\end{eqnarray}

M-XXXX equation
\begin{eqnarray} 
2iS_t=[S,S_{xxxx}]+2\left\{((1+\mu)\vec{S}^2_x-u+m)[S,S_x]\right\}_x,\\
\rho u_{tt}=\nu^2_0u_{xx}++\lambda(\vec{S}^2_x)_{xx}.
\end{eqnarray}

M-XXXXI equation:
\begin{eqnarray} 
2iS_t=\{(\mu \vec S^2_x - u +m)[S,S_x]\}_x,\\
u_t+u_x +\alpha(u^2)_x+\beta u_{xxx}+\lambda (\vec S^2_x)_{x} = 0.
\end{eqnarray}

M-XXXXII equation:
\begin{eqnarray} 
2iS_t=\{(\mu \vec S^2_x - u +m)[S,S_x]\}_x,\\
u_t+u_x +\lambda (\vec S^2_x)_x = 0.
\end{eqnarray}

M-XXXXIII equation:
\begin{eqnarray} 
2iS_t=\{(\mu \vec S^2_x - u +m)[S,S_x]\}_x,\\
\rho u _{tt}=\nu^2_0 u_{xx}+\alpha (u^2)_{xx}+\beta u_{xxxx}+ \lambda
(\vec S^2_x)_{xx}.
\end{eqnarray}

M-XXXXIV equation:
\begin{eqnarray} 
2iS_t=\{(\mu \vec S^2_x - u +m)[S,S_x]\}_x,\\
\rho u _{tt}=\nu^2_0 u_{xx}+\lambda(\vec S^2_x)_{xx}.
\end{eqnarray}

M-XXXXV equation:
\begin{eqnarray} 
2iS_t=[S,S_{xx}]+(uS_3+h)[S,\sigma_3],\\
u_t+u_x+\alpha(u^2)_x+\beta u_{xxx}+\lambda(S^2_3)_x=0.
\end{eqnarray}

M-XXXXVI equation:
\begin{eqnarray} 
2iS_t=[S,S_{xx}]+(uS_3+h)[S,\sigma_3],\\
u_t+u_x+\lambda(S^2_3)_x=0.
\end{eqnarray}

M-XXXXVII equation:
\begin{eqnarray} 
2iS_t=[S,S_{xx}]+(uS_3+h)[S,\sigma_3],\\
\rho u_{tt}=\nu^2_0 u_{xx}+\alpha(u^2)_{xx}+\beta u_{xxxx}+\lambda (S^2_3)_{xx}.
\end{eqnarray}

M-XXXXVIII equation:
\begin{eqnarray} 
2iS_t=[S,S_{xx}]+(uS_3+h)[S,\sigma_3],\\
\rho u_{tt}=\nu^2_0 u_{xx}+\lambda(S^2_3)_{xx}.
\end{eqnarray}

M-XXXXIX equation:
\begin{eqnarray} 
2iS_t=[S,S_{xx}]+(u+h)[S,\sigma_3],\\
u_t+u_x+\alpha(u^2)_x+\beta u_{xxx}+\lambda(S_3)_x=0.
\end{eqnarray}

M-XL equation:
\begin{eqnarray} 
2iS_t=[S,S_{xxxx}]+2\left\{((1+\mu)\vec{S}^2_x-u+m)[S,S_x]\right\}_x,\\
\rho u_{tt}=\nu^2_0u_{xx}+\lambda(\vec{S}^2_x)_{xx}.
\end{eqnarray}

M-XLI equation:
\begin{eqnarray} 
2iS_t=\left\{(\mu\vec{S}^2_x-u+m)[S,S_x]\right\}_x,\\
u_{t}+u_{x}+\alpha(u^2)_{x}+\beta u_{xxx}+\lambda(\vec{S}^2_x)_x=0.
\end{eqnarray}

M-XLII equation:
\begin{eqnarray} 
2iS_t=\left\{(\mu\vec{S}^2_x-u+m)[S,S_x]\right\}_x,\\
u_{t}+u_{x}+\lambda(\vec{S}^2_x)_x=0.
\end{eqnarray}

M-XLIII equation:
\begin{eqnarray} 
2iS_t=\left\{(\mu\vec{S}^2_x-u+m)[S,S_x]\right\}_x,\\
\rho u_{tt}=\nu^2_0u_{xx}+\alpha(u^2)_{xx}+\beta u_{xxxx}+\lambda(\vec{S}^2_x)_{xx}.
\end{eqnarray}

M-XLIV equation:
\begin{eqnarray} 
2iS_t=\left\{(\mu\vec{S}^2_x-u+m)[S,S_x]\right\}_x,\\
\rho u_{tt}=\nu^2_0u_{xx}+\lambda(\vec{S}^2_x)_{xx}.
\end{eqnarray}

M-XLV equation:
\begin{eqnarray} 
2iS_t=[S,S_{xx}]+(uS_3+h)[S,\sigma_3],\\
u_{t}+u_{x}+\alpha(u^2)_{x}+\beta u_{xxx}+\lambda({S}^2_3)_{x}=0.
\end{eqnarray}

M-XLVI equation:
\begin{eqnarray} 
2iS_t=[S,S_{xx}]+(uS_3+h)[S,\sigma_3],\\
u_{t}+u_{x}+\lambda({S}^2_3)_{x}=0.
\end{eqnarray}

M-XLVII equation:
\begin{eqnarray} 
2iS_t=[S,S_{xx}]+(uS_3+h)[S,\sigma_3],\\
\rho u_{tt}=\nu^2_0u_{xx}+\alpha(u^2)_{xx}+\beta u_{xxxx}+\lambda({S}^2_3)_{xx}.
\end{eqnarray}

M-XLVIII equation:
\begin{eqnarray} 
2iS_t=[S,S_{xx}]+(uS_3+h)[S,\sigma_3],\\
\rho u_{tt}=\nu^2_0u_{xx}+\lambda({S}^2_3)_{xx}.
\end{eqnarray}

M-XLVIX equation:
\begin{eqnarray}
2iS_t=[S,S_{xx}]+(u+h)[S,\sigma_3],\\
u_{t}+u_{x}+\alpha(u^2)_{x}+\beta u_{xxx}+\lambda({S}_3)_{x}=0.
\end{eqnarray}

M-L equation:
\begin{eqnarray} 
2iS_t=[S,S_{xx}]+(u+h)[S,\sigma_3],\\
u_{t}+u_{x}+\lambda({S}_3)_{x}=0.
\end{eqnarray}

M-LI equation:
\begin{eqnarray} 
2iS_t=[S,S_{xx}]+(u+h)[S,\sigma_3],\\
\rho u_{tt}=\nu^2_0u_{xx}+\alpha(u^2)_{xx}+\beta u_{xxxx}+\lambda({S}_3)_{xx}.
\end{eqnarray}

M-LII equation:
\begin{eqnarray} 
2iS_t=[S,S_{xx}]+(u+h)[S,\sigma_3],\\
\rho u_{tt}=\nu^2_0u_{xx}+\lambda({S}_3)_{xx}.
\end{eqnarray}

%%%%%%%%%%%%%%%%%%%%%%%%%%%%%%%%
\section{Conclusions}
%%%%%%%%%%%%%%%%%%%%%%%%%%%%%%%%
  Nonlinear models describing interactions of  long and short (LS) waves are given by the Yajima-Oikawa type equations. These long wave - short wave interaction  models  have been derived and proposed with various motivations, which mainly come from fluid and plasma physics. It is well known that in these  long wave-short wave equations is that a long wave always arises
as generated by short waves.  In this paper, we study some of integrable LS  models, namely, the Yajima-Oikawa equation, the Newell equation, the Ma equation, the Geng-Li equation and etc. Any integrable equations admitting the Lax representations, generally speaking, are  gauge equivalent to some integrable generalized HFE. In this context, it is interesting to find, the gauge equivalent counterparts of the above mentioned integrable LS models.  In this paper,  the gauge equivalent counterparts of integrable LS models (equations) are found. In fact, these gauge equivalents of the LS equations are integrable generalized Heisenberg ferromagnet equations  with self-consistent potentials (HFESCP). The associated Lax representations of these HFESCP are given.   We also presented  several spin-phonon equations which describe nonlinear interactions of spin and lattice subsystems in ferromagnetic materials. 

%%%%%%%%%%%%%%%%%%%%%%%%%%%%%%%%%%%%%%%%%%%%%%%%%%%%%
\section*{Acknowledgements}  This work was supported  by  the Ministry of Education  and Science of Kazakhstan, Grant AP08856912.


\begin{thebibliography} {99}
\bibitem{yoe} N. Yajima and M. Oikawa, “Formation and interaction of sonic-Langmuir solitons: Inverse
scattering method,” Prog. Theor. Phys., vol. 56, pp. 1719–1739, 1976.

\bibitem{newell} A. C. Newell. \textit{Long waves-short waves; a solvable model},  SIAM J. Appl. Math., {\bf  35},
650–664 (1978)
\bibitem{ma} Y. C. Ma, “The complete solution of the long-wave short-wave resonance equations,” Stud.
Appl. Math., vol. 59, pp. 201–221, 1978.
\item Zakharov V E 1980 { \it Solitons} ed R K Bullough and P J Caudrey
(Berlin: Sprinder)
\bibitem{melnikov} V.K. Mel'nikov. \textit{On Equations Integrable by the Inverse Scattering
Method}, JINR, preprint P2-85-958, Dubna, 1985
\bibitem{geng} X. Geng, R. Li. \textit{On a vector modified Yajima–Oikawa
long-wave–short-wave equation}, Mathematics, {\bf 7},  958 (2019)
\bibitem{1302.7006} B. Huard, V. Novikov. \textit{On Classification of Integrable Davey-Stewartson Type Equations}, J. Phys. A: Math. Theor., {\bf 46}, 275202 (2013). [arXiv:1302.7006]

\bibitem{maccari1} A. Maccari. \textit{The Kadomtsev-Petviashvili equation as a source
of integrable model equations}, J. Math. Phys., {\bf  37}, 6207 (1996)
\bibitem{maccari2} A. Maccari. \textit{A new integrable Davey-Stewartson-type equation}, 
J. Math. Phys., {\bf  40}, 3971 (1999)
\bibitem{2109.04296} Marcos Caso-Huerta, Antonio Degasperis, Sara Lombardo, Matteo Sommacal. \textit{A new integrable model of long wave-short wave interaction and linear stability spectra}, Proc. R. Soc. A., {\bf  477}, 20210408 (2021). [arXiv:2109.04296]
\bibitem{faquir}  M. Faquir, M. A. Manna, A. Neveu. \textit{An integrable equation governing short waves in a long-wave model}, Proc. R. Soc. A. {\bf 463}, 1939–1954 (2007)
\bibitem{prykarpatsky1} D. Blackmore, Y. Prykarpatsky, 
J. Golenia, A. Prykarpatsky. \textit{Hidden Symmetries of Lax Integrable Nonlinear Systems}, Applied Mathematics, 2013, {\bf 4}, 
95-116 (2013)

\bibitem{ling} L. Ling and Q. P. Liu. \textit{A long waves-short waves model: Darboux transformation and soliton
solutions}, J. Math. Phys., {\bf 52},  053513 (2011)


\bibitem{ma} R. Li and X. Geng, “On a vector long wave-short wave-type model,” Stud. Appl. Math., vol. 144,
pp. 164–184, 2020.
\bibitem{ma} R. Li and X. Geng, “A matrix Yajima Oikawa long-wave-short-wave resonance equation, Darboux transformations and rogue wave solutions,” Commun. Nonlinear Sci., vol. 90, p. 105408,
2020.
\bibitem{ustinov} N.V. Ustinov. \textit{Appearing (disappearing) lumps and rogue lumps of the
two-dimensional vector Yajima–Oikawa system}, [arXiv:1911.10583]
\bibitem{ishimori} Y. Ishimori. \textit{Multi-vortex solutions of a two-dimensional nonlinear wave equation}, Prog. Theor. Phys., {\bf 72}, 33-37 (1984)
\bibitem{s1} Chen Chi, Zhou Zi-Xiang. \textit{Darboux Transformation and Exact Solutions of the Myrzakulov-I Equation}, Chin. Phys. Lett., {\bf 26}, N8, 080504 (2009)
\bibitem{s2} Chen Hai, Zhou Zi-Xiang. \textit{Darboux Transformation with a Double Spectral Parameter for the Myrzakulov-I
Equation}, Chin. Phys. Lett., {\bf 31}, N12, 120504 (2014)
\bibitem{s3} Hai Chen,  Zi-Xiang Zhou. \textit{Global explicit solutions with n double spectral parameters
for the Myrzakulov-I equation}, Modern Physics Letters B, {\bf 30}, N29, 1650358 (2016)
\bibitem{s4} Hai-Rong Wang, Rui Guo. \textit{Soliton, breather and rogue wave solutions for the Myrzakulov-Lakshmanan-IV equation}, Optik, 
{\bf 242}, 166353 (2021)
 \bibitem{rm1}	R. Myrzakulov, A. Danlybaeva and G. Nugmanova. 	Theor. and Math. Phys., V.118, 13, P. 441-451 (1999).
\bibitem{RM3} R. Myrzakulov, G. Mamyrbekova, G. Nugmanova, M. Lakshmanan. Symmetry, {\bf 7}(3), 1352-1375 (2015).
\bibitem{RM4} Z. S. Yersultanova, M. Zhassybayeva, K. Yesmakhanova, G. Nugmanova,  R. Myrzakulov. \textit{Darboux transformation and exact solutions of the integrable Heisenberg ferromagnetic equation with self-consistent potentials},  Int. Jour. Geom. Meth. Mod. Phys., {\bf 13}, N1, 1550134 (2016).
\bibitem{RM20} R. Myrzakulov, G. N. Nugmanova,  R. N. Syzdykova. \textit{Gauge equivalence between (2+1)-dimensional continuous Heisenberg ferromagnetic models and nonlinear Schr\"{o}dinger-type equations},  J. Phys. A: Math. Gen., {\bf  31}, N47,  9535-9545 (1998)
\bibitem{RM5} R. Myrzakulov, S. Vijayalakshmi, R. Syzdykova, M. Lakshmanan, \textit{On the simplest (2+1) dimensional integrable spin systems and their equivalent nonlinear Schrödinger equations}, J. Math. Phys., {\bf 39},   2122-2139 (1998).
\bibitem{makhankov} V.G. Makhankov, R. Myrzakulov. \textit{$\sigma$-Model Representation of the Yajima-Oikawa Equation System}, Preprint P5-84-719, JINR, Dubna, 1984 
\bibitem{RM0} V. G. Makhankov, R. Myrzakulov, O. K. Pashaev. \textit{Gauge Equivalence, Supersymmetry and Classical Solutions of the ospu(1,1/1) Heisenberg Model and the Nonlinear Schr\"{o}dinger Equation}, Letters in Mathematical Physics, {\bf  16}, 83-92 (1988). 
\bibitem{RM1} R. Myrzakulov, S. Vijayalakshmi, G.N. Nugmanova, M. Lakshmanan. \textit{A (2+1)-dimensional integrable spin model:
Geometrical and gauge equivalent counterpart, solitons
and localized coherent structures}, Phys. Lett. A, {\bf  233},  391  (1997). [arXiv:solv-int/9704005
\bibitem{rm51} R. Myrzakulov, O. K. Pashaev,  Kh. T. Kholmurodov. \textit{Particle-Like Excitations in Many Component Magnon-Phonon Systems}, Physica Cripta, {\bf 33}, N4, 378 (1986)
\bibitem{RM6} R. Myrzakulov, M. Lakshmanan, S. Vijayalakshmi, A. Danlybaeva, J. Math. Phys., {\bf 39},  3765-3771 (1998).

\bibitem{RM7} R. Myrzakulov, G. Nugmanova, R. Syzdykova, Journal of Physics A: Mathematical \& Theoretical, {\bf 31} 147, 9535-9545 (1998).

\bibitem{RM8} R. Myrzakulov, M. Daniel, R. Amuda,  Physica A., {\bf 234}, 13-4, 715-724 (1997).

\bibitem{RM9} S.C. Anco, R. Myrzakulov, Journal of Geometry and Physics, v.60, 1576-1603 (2010).

\bibitem{RM10} R. Myrzakulov, G. K. Mamyrbekova, G. N. Nugmanova, K. Yesmakhanova,  M. Lakshmanan.   Physics Letters A, {\bf 378}, 30-31, 2118-2123 (2014).


\bibitem{nevin1} Nevin Ertug G\"{u}rb\"{u}z, R. Myrzakulov, Z. Myrzakulova. \textit{Three anholonomy densities for three formulations with anholonomic coordinates with hybrid frame in $R^{3}_{1}$ }, Optik, 169161 (2022)
\bibitem{kemal} K. Eren, K.  Yesmakhanova, S. Ersoy, R. Myrzakulov. \textit{Involute Evolute Curve Family Induced by the  Coupled Dispersionless Equations}
(in preparation)

\bibitem{RM11} L. Martina,  Kur. Myrzakul,  R. Myrzakulov, G. Soliani, Journal of Mathematical  Physics, V.42, 13, P.1397-1417 (2001).
 \bibitem{RM22} A.  Myrzakul, G. Nugmanova, N.  Serikbayev, R. Myrzakulov. \textit{Surfaces and Curves Induced by Nonlinear Schr\"{o}dinger-Type Equations and Their Spin Systems}, Symmetry, {\bf 13}, N10, 1827 (2021) 
\bibitem{RM12} A. Myrzakul and R. Myrzakulov. \textit{Integrable geometric flows of interacting curves/surfaces, multilayer spin systems and the vector nonlinear Schrödinger equation},  International Journal of Geometric Methods in Modern Physics,  {\bf 14}, N10, 1750136 (2017)]
\bibitem{rm21} A. Myrzakul and R. Myrzakulov. \textit{Integrable motion of two interacting curves, spin systems and the Manakov system}, International Journal of Geometric Methods in Modern Physics,  {\bf 14}, N7, 1750115 (2017)
\bibitem{laksh} M. Lakshmanan. Phys. Lett. A, {\bf 64}, 53-54 (1977) 
\bibitem{zt1979} V. E. Zakharov, L. A. Takhtajan,  Theor. Math. Phys., {\bf 38},  17-23  (1979)
\bibitem{zhanna1} Z. Sagidullayeva, K. Yesmakhanova,  N.  Serikbayev, G. Nugmanova, K.  Yerzhanov, R. Myrzakulov \textit{Integrable generalized Heisenberg  ferromagnet   equations in 1+1 dimensions:   reductions and gauge equivalence}, [arXiv:2205.02073]
\bibitem{2206.05348} Z. Sagidullayeva, G.   Nugmanova, R.  Myrzakulov and   N. Serikbayev. \textit{Integrable Kuralay equations: geometry, solutions and generalizations}, Symmetry, {\bf 14}, N7, 1374 (2022).  [arXiv:2206.05348]
\bibitem{z1}  Z. Sagidullayeva, K. Yesmakhanova, G. Nugmanova,  R. Myrzakulov \textit{Soliton solutions of the Kuralay equation via Hirota bilinear method}. Book of Abstracts, 6th NMMP-2022, FAMU, Tallahassee, USA, June 17-19, 2022

\bibitem{z3} Z. Myrzakulova, G. Nugmanova, K. Yesmakhanova, R. Myrzakulov. \textit{Integrable motion of anisotropic space curves and surfaces induced by the Landau-Lifshitz equation}, [arXiv:2202.00748]



\end{thebibliography}
\end{document}